# Experimental measurement of the intrinsic excitonic wavefunction

Michael K. L. Man*(1), Julien Madéo*(1), Chakradhar Sahoo(1,2), Kaichen Xie(3), Marshall Campbell(4), Vivek Pareek(1), Arka Karmakar(1), E Laine Wong(1)†, Abdullah Al-Mahboob(1), Nicholas S. Chan(1), David R. Bacon(1), Xing Zhu(1), Mohamed Abdelrasoul(1), Xiaoquin Li(4), Tony F. Heinz(5,6), Felipe H. da Jornada(7), Ting Cao(3,5), Keshav M. Dani(1)‡

(1)     Femtosecond Spectroscopy Unit, Okinawa Institute of Science and Technology, 1919-1 Tancha, Onna-son, Okinawa, Japan 904-0495

(2)     School of Physics, University of Hyderabad, Gachibowli, Hyderabad-500046, Telangana, India.

(3)     Department of Materials Science and Engineering, University of Washington, Seattle, WA 98195, USA.

(4)     Physics Department, Center for Complex Quantum System, The University of Texas at Austin, Austin, TX 78712, USA.

(5)     Department of Applied Physics, Stanford University, Stanford, CA 94305, USA.

(6)     SLAC National Accelerator Laboratory, Menlo Park, CA 94720, USA.

(7)     Department of Materials Science and Engineering, Stanford University, Stanford, CA 94305, USA.

*These authors contributed equally to this work.

†Present address: Center for Nano Science & Technology, Italian Institute of Technology, 20133 Milan, Italy.

‡Corresponding author. Email: KMDani@oist.jp

**Abstract**

An exciton – a two-body composite quasiparticle formed of an electron and hole – is a fundamental optical excitation in condensed-matter systems. Since its discovery nearly a century ago, a measurement of the excitonic wavefunction has remained beyond experimental reach. Here, we directly image the excitonic wavefunction in reciprocal space by measuring the momentum distribution of electrons photoemitted from excitons in monolayer $WSe_2$. By transforming to real space, we obtain a visual of the distribution of the electron around the hole in an exciton. Further, by also resolving the energy coordinate, we confirm the elusive theoretical prediction that the photoemitted electron exhibits an inverted energy-momentum dispersion relationship reflecting the valence band where the partner hole remains, rather than that of conduction-band states of the electron.

**Main Text**

In the 1930s, pioneering work by Frenkel [1], Wannier [2] and others [3,4] elucidated the presence of optically excited states at energies lying in the forbidden bandgap region of insulators and semiconductors. These states, called excitons, could not be described within a single-particle picture, but arose as correlated bound states between a photoexcited electron in the conduction band and the hole left behind in the valence band. As neutral composite quasiparticles of oppositely charged fermions, excitons exhibit different quantum statistics and respond differently to external fields compared to their constituent free carriers. Over the ensuing century, these differences have had inevitable consequences for various condensed matter systems, giving rise to phenomena such as Bose-Einstein condensation [5–7] and excitonic insulators [8], and impacting the performance of photovoltaics [9,10], light-emitted diodes [11,12] and other opto-electronic devices [13].

As in the case of other two-body systems, the fundamental theoretical description of excitons is naturally formulated in terms of the relative electron-hole coordinates, particularly for the Wannier-type excitons, where electron-hole separation extends over many lattice sites [14,15]. Such descriptions, constituting the excitonic wavefunction in real space – analogous to the well-known wavefunctions describing the hydrogen atom – directly define the exciton's size and shape. The excitonic wavefunction can also be described in momentum space, which dictates its ability to interact with light, phonons, plasmons and other quasiparticles through momentum conservation [16]. However, since their discovery nearly a century ago, measuring the excitonic wavefunction, whether in real- or reciprocal-space, has not been possible [17]. This is in part due to the small binding energy and finite lifetimes of excitons in most semiconductor systems, but also due to limitations in available experimental techniques. Optical techniques provide precise spectroscopic information about the exciton, but do not access the momentum coordinates of the constituent electrons and holes, while techniques like STM and TEM with very high spatial resolution cannot measure the relative distance between the delocalized electrons and holes in extended systems.

Over the past decade, the discovery of robust excitons in two-dimensional systems [18–20] and advances in space-, time- and angle-resolved photoemission spectroscopy (ARPES) techniques [21–24] have created new opportunities in this regard. Analogous to collider experiments of high energy physics, theoretical studies from the past few years have proposed using a TR-ARPES framework to dissociate the exciton with a high-energy XUV photon, and photoemit its constituent electron [17,25,26]. Under the right conditions, the momentum of the photoemitted electron corresponds to the relative electron-hole momentum in the exciton, and its measured distribution directly reflects the excitonic wavefunction in reciprocal space [25,27,28]. Very recently, a time-resolved ARPES experiment on monolayer $WSe_2$ – a prototypical 2D semiconductor, validated this approach by resolving excitons with different relative electron-hole momenta, namely, bright and momentum dark excitons [29]. This successful demonstration immediately raises the tantalizing possibility of imaging the excitonic wavefunction.

Here, we directly image the exciton wavefunction in momentum space for the K-valley exciton in monolayer $WSe_2$. The corresponding real-space wavefunction, obtained by Fourier transformation, describes the distribution of the electron relative to the hole in the exciton, revealing a radius of 1.4 nm that extends over many lattice sites. By energy-resolving the momentum distribution, we also observe the decades old prediction [23,28,30] of the anomalous dispersion effect that has been experimentally elusive under different measurement conditions [30,31]: The photoemitted electron mimics the *negative* energy-momentum dispersion of the valence band of the hole, to which it was previously bound in the exciton, rather than the positive dispersion of its own conduction band.

In our study, we examined an exfoliated monolayer (ML) of $WSe_2$ transferred onto a thin hBN buffer layer supported by a Si substrate (Fig. 1A). The insulating hBN layer provides a clean, flat substrate, while also preventing the quenching of the exciton [32] (fig. S1). The valence band structure of our sample prior to photoexcitation was measured using a ultrashort (160 fs) probe pulses at a photon energy of 21.7 eV in XUV to photoemit electrons from the sample. Figure S2 confirms key features of the band structure of monolayer $WSe_2$ prior to photoexcitation – the valence band maximum (VBM) at the K-valley, a splitting of the VB at the K-valley of 0.47 eV due to spin orbit coupling, and the VBM at the $\Gamma$-valley situated 0.57 eV lower in energy than the K-valley. A comparison to a theoretical, first-principles GW calculation of the VB shows very good agreement.

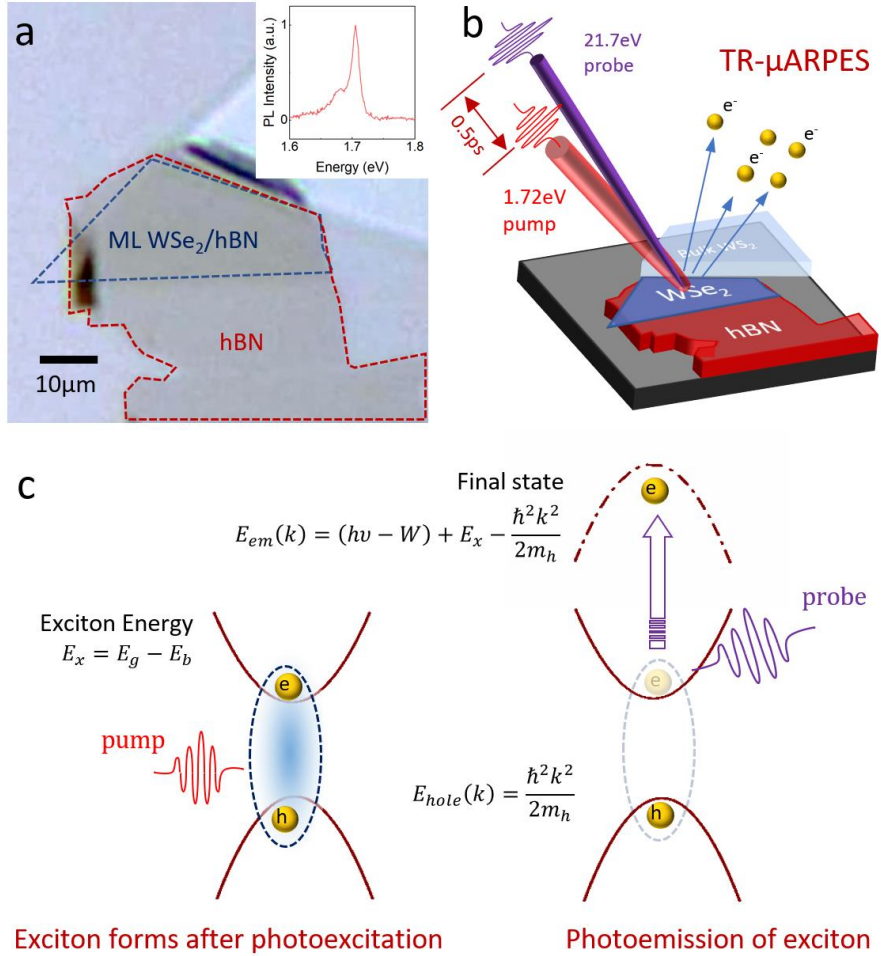

**Fig 1. TR-µ-ARPES experimental scheme and the anomalous dispersion of the photoemitted electron.** (a) Optical image of the WSe$_2$/hBN/Si heterostructure, with inset showing photoluminescence spectrum of the monolayer taken at 90 K. (b) In our TR-µ-ARPES setup, we resonantly create excitons in the WSe$_2$ ML using a low-intensity ultrafast pump pulse, and photoemit electrons from the excitons using a time-delayed XUV probe pulse. (c) As a result of the photoemission process, due to energy and momentum conservation, the energy-momentum dispersion relationship of the photoemitted electron resembles its partner-hole's valence band. (see the supplementary materials)

To measure the excitonic states, we resonantly photoexcited the A-exciton transition of the WSe$_2$ monolayer at a photon energy of 1.72 eV with a 175 fs linearly polarized pump pulse (Fig 1b and fig. S1). This created direct excitons with holes and electrons in the K-valley. Subsequently, these excitons scattered into other excitonic states including spin-dark K-K excitons, indirect-bandgap K-K' excitons and Q-K excitons [29], where the two letters denote the valleys of the electron and hole, respectively. To dissociate the excitons and photoemit the constituent electrons, we employed a time-delayed XUV probe pulse as above. As expected [25–29], we observed the below bandgap signal corresponding to electrons photoemitted from the K-valley excitonic states (fig. S2). Further information about the experimental setup for probing the WSe$_2$ ML band structure before and after photoexcitation, the ensuing exciton dynamics, and the attribution of the signal to the K-valley excitons have been previously reported [29].

We now turn our attention to understanding the conditions under which the above measurement provides access to the excitonic wavefunction. For this, let us consider an exciton in the low-density limit at zero temperature. For such an exciton $X$, the wavefunction $|\phi_X\rangle = \sum_{vc\mathbf{k}} A^X_{vc\mathbf{k}} c^\dagger_{c\mathbf{k}} c_{v\mathbf{k}} |0\rangle$ describes the electron-hole bound state, where $|0\rangle$ is the system ground state and $c_{i\mathbf{k}}$ destroys an electron with band index $i$ and wavevector $\mathbf{k}$. The factors $A^X_{vc\mathbf{k}}$ are then expansion coefficients of the exciton in terms of free electron-hole interband pairs, which can be interpreted as the envelope function of the Wannier exciton wavefunction in reciprocal space [14,27]. Thus, in the Wannier limit, where the wavefunctions of the constituent electrons are similar in nature as a function of $\mathbf{k}$, the probability of the XUV probe pulse photoemitting a constituent electron from the exciton with momentum $\mathbf{k}$ is proportional to $|A_{\mathbf{k}}|^2$. In other words, under low-temperature and low-density quasi-equilibrium conditions, the momentum-resolved intensity distribution of the photoemitted electrons in $k$-space directly images the modulus squared of the excitonic envelope function in reciprocal space, which we will refer to as simply the wavefunction squared.

In our experiment, as we lower the exciton density by lowering the optical pump pulse intensity, we see a rapid reduction in the width of the measured $k$-space distribution, as broadening effects due to many-exciton interactions decrease (fig. S3). For densities <$10^{12}$ cm$^{-2}$, one expects minimal exciton-exciton interactions since the average inter-excitonic distance exceeds the expected (few nanometer) size of the exciton [33]. Thus, the width of the ARPES peak approaches the intrinsic width of the exciton in $k$-space. Figure 2a shows the measured excitonic wavefunction-squared at a total exciton density of ~1 x $10^{11}$ cm$^{-2}$, as recorded at a time-delay of 0.5 ps. We find that for the resonant, low-excitation conditions, the measured $k$-space distribution is largely unchanged as a function of time delay (fig. S4), suggesting that quasi-equilibrium conditions are rapidly achieved within our few hundred femtosecond temporal resolution. A calculation utilizing first-principles GW-BSE techniques reproduces the measured excitonic wavefunction in the full two-dimensional reciprocal space very well (Fig. 2b, d, e). From our measurements, we obtain a root mean squared (RMS) radius in *k*-space of 0.072 1/Å, which corresponds to an RMS radius of 1.4 nm in real space (Fig. 2c) (see the supplementary materials for details), in excellent agreement with our theory (Fig. 2f,g). Previous magneto-optical studies on $WSe_2$ have reported a radius of 1.7 nm (RMS) for encapsulated samples, where a slightly larger radius is expected because of the increased dielectric screening from the encapsulating hBN layers [33].

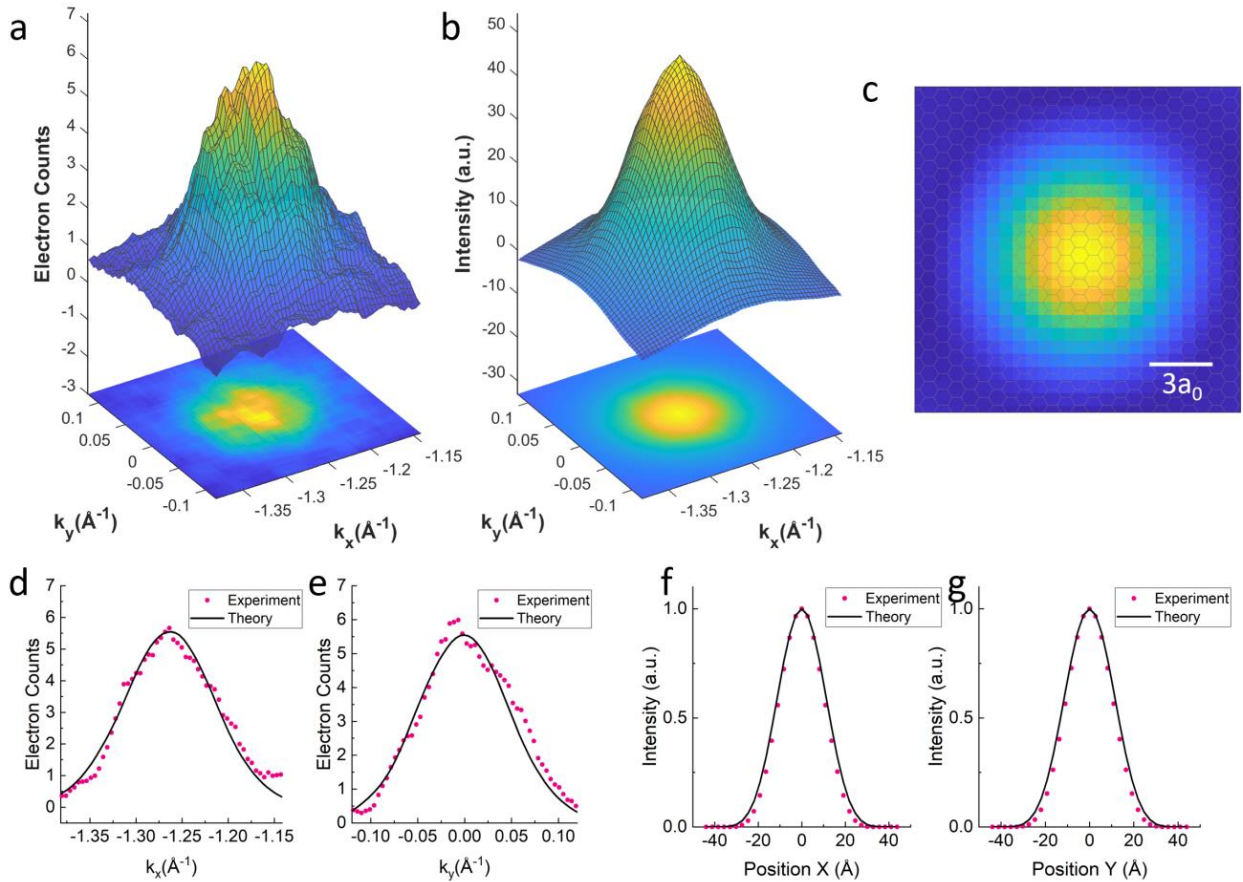

**Fig 2**. **Excitonic wavefunction-squared of the K-valley exciton.** (a) experiment and (b) theory in momentum space. (c) Wavefunction-squared in real space obtained by Fourier transform of the experimental data. The overlay of the $WSe_2$ lattice shows the relative electron-hole distances extend over many lattice sites – the hallmark of a Wannier-type exciton. (d - f) shows profile comparisons of experiment and theory in two orthogonal directions in momentum (d, e) and real space (f, g).

Besides the momentum resolution of the wavefunction, our experiments also allow us to resolve the energy of the photoemitted electron (Fig. 3) and thus measure the dispersion relation of electrons originating from excitons. Under our low temperature and low density conditions, we have the opportunity to evaluate the predicted anomalous dispersion [27,28,30], a signature of excitons that has been elusive in experimental studies performed under different conditions [30,31]. Based on momentum and energy conservation, one can show that electrons bound to holes in excitons, upon being photoemitted, will exhibit a *negative* energy-momentum dispersion relationship that resembles the valence band of the hole (see the supplementary materials for a detailed description of the phenomenon).

In Fig. 3b, we plot the energy-momentum distribution of the photoemitted electron versus k along the K-Γ direction of the BZ. For each k, we see a distribution in the measured photoelectron energies (Fig. 3C), which is largely due to the inhomogeneous broadening in the photoemission signal from our sample. Despite this energy distribution, one clearly sees the negative dispersion exhibited by the photoemitted electron. To analyze this further, we extract the peak energy of the distribution for every point in the 2D *k*-space (Fig. 3a – yellow surface, see the supplementary materials for details). A similar analysis is also performed for the valence band, thus providing its dispersion over the 2D k-space as well (Fig. 3a – blue surface). Together, they exhibit a striking visual of the negative curvature of the VB as well as the sub-gap excitonic signal. We also note that Fig. 3a provides a remarkable illustration of how an ARPES-based measurement encodes a correlated two-particle state in a single-particle band structure. In Fig. 3B, we plot the peak energy versus *k* for the exciton (magenta line) and the valence band (dashed yellow), showing the comparable negative curvatures of the two signals in an overlay (the valence band is displaced in energy for an easier comparison). We expect any discrepancy to reflect experimental uncertainty, inhomogeneous broadening in the photoemission signal of the sample, and finite temperature effects.

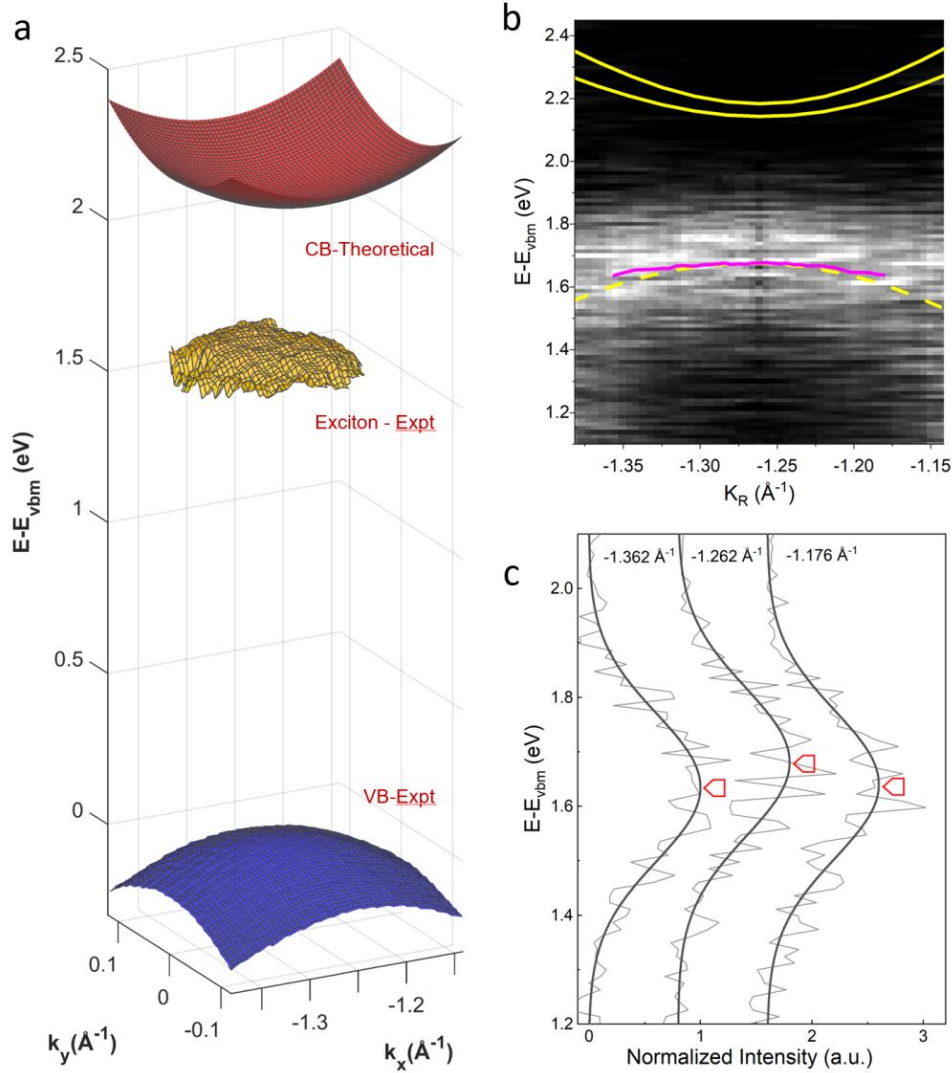

**Fig 3. Anomalous negative dispersion of the electron photoemitted from an exciton.** (a) Measured energy-momentum dispersion relationship over the two-dimensional k-space for the electron photoemitted from the exciton and from the valence band. Strikingly, the exciton signal shows a negative dispersion resembling the valence band. A theory calculation for the dispersion of the conduction band (CB) is also shown for comparison. The dispersion relationship for the exciton (yellow) and VB (blue) are obtained by fitting the experimental Energy Distribution Curves (EDC) and extracting their peak values $E_X(\mathbf{k})$ and $E_{VB}(\mathbf{k})$ at each momentum. (see the supplementary materials) (b) Measured energy-momentum distribution of the electrons photoemitted from an exciton, along a one-dimensional cut in k-space centered at the K-valley. For each value of $k_\parallel$, the intensity versus energy is normalized to its maximum value. The yellow solid curves are the theoretical spin-split conduction bands. The dashed yellow and magenta curves correspond to $E_{VB}(\mathbf{k})$ and $E_X(\mathbf{k})$, with $E_{VB}(\mathbf{k})$ off-set in its energy position for easy comparison to $E_X(\mathbf{k})$. (c) EDCs (light gray) and Gaussian fitting (black) for three representative values of $k_\parallel$ from panel (B), with red △ symbol marking the peak positions of the fitted curve.

Our experiments of photo-dissociating the exciton and measuring the photoemitted electron distribution in energy and momentum spaces allows for unprecedented information about excitons. In the immediate future, one could image interesting excitonic species in monolayer TMDs that are not otherwise easily studied by light, such as the momentum- or spin-forbidden dark excitons. Similar studies in other interesting materials systems, such as the TMD hetero- and homo-bilayers, 2D perovskites, could provide a rich understanding of exciton physics therein. Furthermore, it is likely that the central idea of dissociating composite quasiparticles and measuring their constituents via TR-ARPES, could be extended to study other optically excited states, such as trions and biexcitons. Finally, our work also raises the intriguing question of whether similar studies from an *ensemble* of excitons could reveal the nature of emerging coherent quantum states as one lowers the temperature or increases the density.

**Acknowledgments**
This work was supported by JSPS KAKENHI Grant number JP17K04995 and in part by funding from the Femtosecond Spectroscopy Unit, Okinawa Institute of Science and Technology Graduate University. We thank OIST engineering support section for their support. This research (theoretical analysis and first-principle calculations) was partially supported by NSF through the University of Washington Materials Research Science and Engineering Center DMR-1719797. Analysis at SLAC was supported by the AMOS program, Chemical Sciences, Geosciences, and Biosciences Division, Basic Energy Sciences, US Department of Energy under Contract DE-AC02-76-SF00515. T.C. acknowledges support from the Micron Foundation and a GLAM postdoctoral fellowship at Stanford. K.X. acknowledges support by the state of Washington through the University of Washington Clean Energy Institute. Computational resources were provided by Hyak at UW, and the Extreme Science and Engineering Discovery Environment (XSEDE), which is supported by National Science Foundation under Grant No. ACI-1053575. The work of M.C. and X. L. at Austin was partially supported by the National Science Foundation through the Center for Dynamics and Control of Materials: an NSF MRSEC under Cooperative Agreement No. DMR-1720595 and the facility supported by the center.

**Contributions**
M.K.L.M., J.M., C.S., E.L.W. V.P., and A.A.M performed the experiments with assistance from D.B., M.M.M.A, and X.Z. M.K.L.M., K.M.D., J.M. and N.C. analyzed the data. M.C. and X. L. prepared the sample. V.P. and A.K. characterized the sample. T.C., F.H.J. and T.F.H. provided theoretical support. K.X. and T.C. performed first principle theoretical calculations with help from F.H.J. K.M.D. and TC conceived the experiment. K.M.D. supervised the project. All authors contributed to discussions and manuscript preparation.

**Competing interests**
J.M., M.K.L.M. and K.M.D. are inventors on a provisional patent application related to this work filed by the Okinawa Institute of Science and Technology School Corporation (US 2020/0333559 A1 published on October 22, 2020). The authors declare no other competing interests.

# Supplementary Materials for

## Experimental measurement of the excitonic wavefunction

Michael K. L. Man, Julien Madéo, Chakradhar Sahoo, Kaichen Xie,
Marshall Campbell, Vivek Pareek, Arka Karmakar, E Laine Wong, Abdullah Al-Mahboob,
Nicholas S. Chan, David R. Bacon, Xing Zhu, Mohamed Abdelrasoul, Xiaoquin Li,
Tony F. Heinz, Felipe H. da Jornada, Ting Cao, Keshav M. Dani

Correspondence to: KMDani@oist.jp

**Materials and Methods**

Sample preparation

WSe$_2$ and hBN were mechanically exfoliated and stacked onto a *n*-doped Si substrate by dry-transfer technique. Size of the monolayer WSe$_2$ is around 40 x 20 µm. The hBN buffer is around 20 nm think, this buffer layer provides a clean and flat support, and it also provide a dielectric environment that prevent quenching of exciton. The WSe$_2$ monolayer was connected to bulk WSe$_2$ that sits directly on top of the Si substrate, this provides a conductive pathway and prevents sample charging. More details about the sample preparation and geometry can be found in [29].

Optical pump and XUV probe

Our optical setup has been described in detail previously [29]. The optical pump and probe are driven by an Ytterbium-doped fiber amplifiers laser system that operates at 1 MHz, providing 230 fs pulses at 1030 nm and pulse energy of 100 µJ. 20 µJ of this pump is used to drive a noncollinear optical parametric amplifier (NOPA), which provides a tunable wavelength from 320 to 2500 nm with 5 nm spectral bandwidth and with pulse energies ranging from 0.1 - 1 µJ. We use the output of this NOPA to photoexcite the sample, incident angle is 68° from surface normal. For the low temperature experiment, we photoexcite the sample resonantly at 1.72 eV (with 5 nm bandwidth) with a photoexcitation density of 1.02 x 10$^{11}$/cm$^2$. Details about estimating the photoexcitation density are provided in supplementary materials.

For the XUV probe, with a BBO crystal, we frequency doubled 70 µJ of 1030 nm from the laser system generating 30 µJ of 515 nm radiation. High harmonic generation is done by focusing the 515 nm beam into a Kr gas jet in vacuum, reaching a power density of 3 x $10^{14}$ W/cm². Out of the different harmonics, 21.7eV photon was selected by a combination of Al and Sn filters. In all the experiments, the probe pulse was p- polarized and incident on the sample at 68° from surface normal. The effective temporal resolution of this pump-probe setup was ~240 fs.

Time-resolved angle resolved photoemission spectroscopy (TR-ARPES)

TR-ARPES was performed in a time-of-flight momentum microscope [29]. Photoelectrons emitted from the sample were collected by an immersion objective lens, providing access to full half-space above the sample surface. A field aperture inserted at the Gaussian image plane of the electron optics was used to selectively allow only electrons emitted from the monolayer region of the sample to pass through. Momentum space (ARPES) image of this selected sample area was obtained by imaging the back focal plane of the objective lens, and energy resolved spectrum was measured by a time-of-flight detector. Energy resolution of the instrument was determined by the drift energy of electrons in the time-of-flight drift tube, in this experiment, the effective energy resolution was set to 30 meV.

First-principles calculations

Density functional calculations (DFT) within the local density approximation (LDA) were performed using the Quantum ESPRESSO package [34]. We used the experimental lattice constant of 3.28 Å in our calculations. The GW [35] calculations were carried out using the BerkeleyGW package [36]. In the calculation of the electron self-energy, the dielectric matrix was constructed with a cutoff energy of 35 Ry. The dielectric matrix and the self-energy were calculated on an $18 \times 18 \times 1$ k-grid. The quasiparticle bandgap was converged to within 0.05 eV. The spin–orbit coupling was included perturbatively within the LDA formalism. The calculations of the excitonic ARPES spectra use ensemble average of photoelectrons emitted from excitons of different momenta, with the exciton population following quasi-equilibrium Boltzmann distribution in the K valley. The exciton energy levels and wavefunction are calculated using the GW-BSE methods implemented in the BerkeleyGW package, with the k-grid sampled in the K-valleys and the grid density equivalent to 14,400 points in the first Brillouin zone [37]. In general, the GW-BSE method is expected to be accurate to within roughly 100 meV for quasiparticle band gaps of semiconductors [38]. Another important consideration is the convergence parameters coming from the number of k points, number of empty bands, and dielectric cut-off which in our case yield an error smaller than 50 meV. The calculated energy levels of K-valley excitons can be further impacted by factors such as the choice of lattice constant and the pseudopotential used in the calculations.

**Supplementary Materials**

Excitonic Resonances

To ensure that we are pumping resonantly at the A-excitons, we measured the photoemission excitation response across different excitation photon energies in TR-ARPES. Figure S1 shows a plot where we measured the integrated photoemission intensity from 1 to 3eV above the VB, capturing the excitonic signal, at different photon energy. At 90 K, we observed strong resonant of our photoemission at 1.72 eV, corresponding to the optically bright A-exciton and matches well with the peak position of the photoluminescence spectrum as shown in Fig.1a. In addition, we observe other resonances at higher photon energy, related to B- and C-excitons in monolayer $WSe_2$.

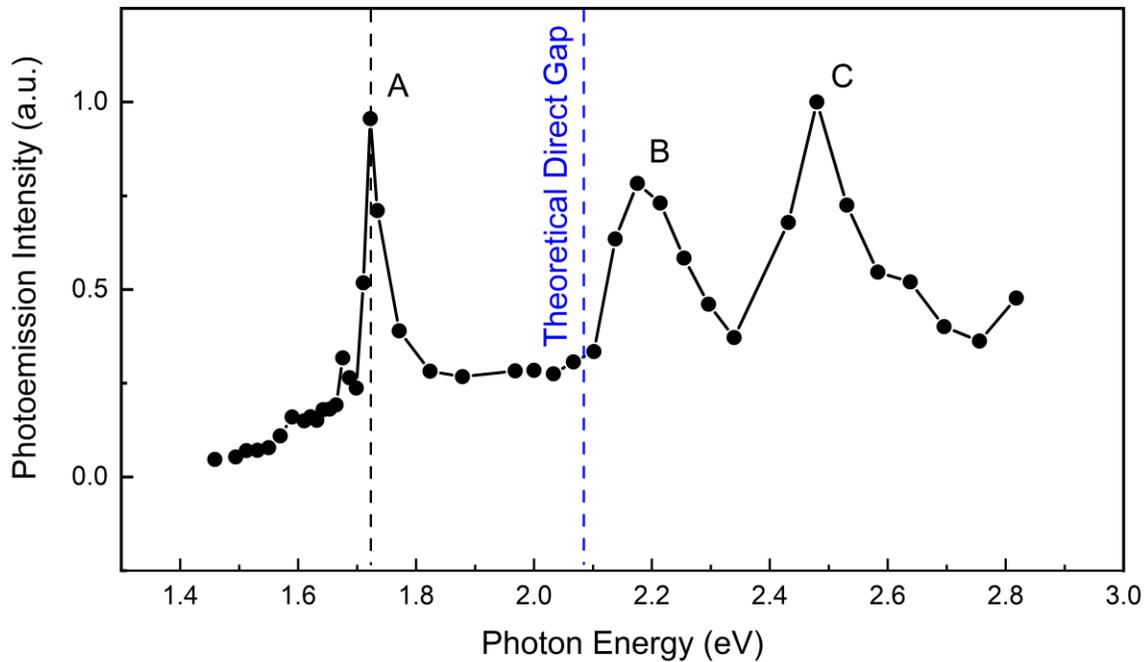

**Fig. S1.** Plot of photoemission intensity versus pump photon energy taken at 90K. Adapted from [29].

Unexcited and excited ARPES data

Our ARPES data obtained by the time-of-flight momentum microscope comes in a form of 3D datacube ($K_x$, $K_y$, E). Fig S2 shows a slide of the ARPES spectra along the K-Γ direction. To improve the signal-to-noise ratio, we averaged signal in a range of about 0.04 Å$^{-1}$ in the direction orthogonal to the cut. To show the exciton clearly, we enhanced the intensity for the exciton in Fig. S4b by 200 times for data 0.5eV above the VBM. Integration time for the no pump and 1.72 eV pump data are 30 mins and 42 hours, respectively. The white lines show the first-principles calculated bandstructure.

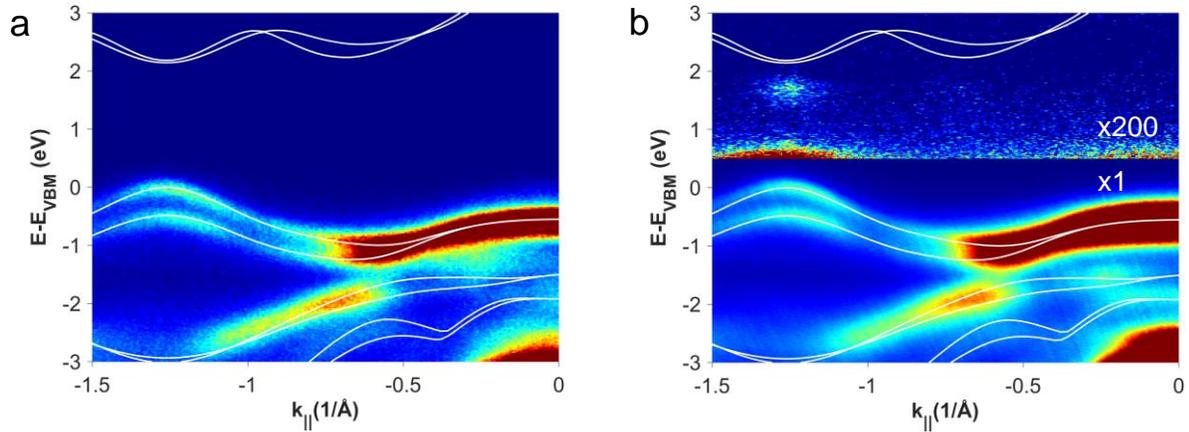

**Fig. S2.** ARPES spectra along the K-Γ direction taken (a) without pump, and (b) with 1.72eV pump. White lines are the calculated quasiparticle band structures.

Intensity dependence of the measured k-space distribution

One expects broadening in the measured distribution of the photoelectron's momentum as a function of exciton density due to various nonlinear and many-exciton effects. As shown in Fig. S6, we observe such broadening at exciton densities $>10^{12}$ cm$^{-2}$, where the average inter-excitonic distance starts to become comparable to the exciton radius. At low enough densities, one can minimize these broadening effects and obtain a width of the k-space distribution that corresponds to the finite size of the exciton.

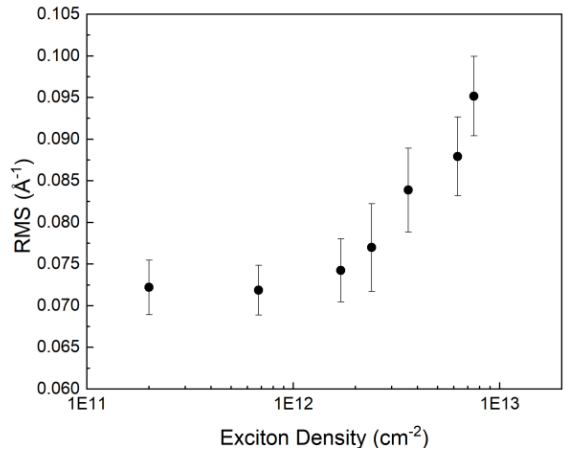

**Fig.S3.** Size of the excitonic wavefunction measured at different exciton densities. Measurement were done at a time-delay of 1ps, at sample temperature of 90K.

Achieving Quasi-equilibrium Conditions

We see the width of k-space distribution remain largely unchanged as a function of pump-probe delay for these very low excitation conditions ($< 10^{12}$ cm$^{-2}$ at 0 ps), indicating the quasi-equilibrium conditions are rapidly achieved within our few-hundred femtosecond time-resolution.

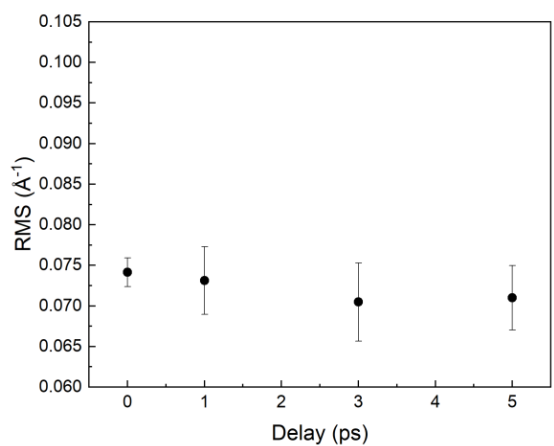

**Fig.S4.** Excitonic wavefunction size versus pump-probe time-delay. Measured at 90K, starting with an initial exciton density is $8.2 \times 10^{11}$ cm$^{-2}$.

Dispersion fitting for VB and Exciton

In figure 3a, we plotted the E(**k**) dispersion of the valence band (VB) and exciton in the two-dimensional momentum space. For VB, we first obtain energy distribution curves (EDC) at different k position in momentum space around the K valley (Fig. S5A). We applied double Gaussian fitting to extract the peak position of the two split bands due to spin-orbit effect (Fig. S3B). Figure S3c shows the dispersion of VB1 along M-K-Γ direction at $k_y = 0$ Å$^{-1}$. By fitting of the 2D dispersion surface up to a radius of 0.07 Å$^{-1}$, we obtained a hole effective mass of -0.41 $m_e$ (bare electron mass) for VB1 at the K-valley.

We obtained the 2D E(**k**) dispersion of the exciton by single Gaussian fitting at each k. Due to the much lower electron counts in the excitons, we apply data averaging before obtaining EDCs. For each k position, we sum up the electron counts within an area of 0.04 x 0.04 Å$^{-2}$ around the k. We are aware that data averaging could affect E(**k**) dispersion, i.e. it could flatten the bands and effectively alter the measured effective mass. We check how this averaging procedure affect the effective mass determination in the VB (Fig S6), and found that without any averaging (when we only take 1 pixels in the image, which corresponds to an area of 0.0045 x 0.0045 Å$^{-2}$), we obtain an effective mass of -0.411 . With averaging, the effective mass only changes to -0.419. It shows that disregard of data averaging, we can robustly obtain the same effective mass.

In Fig3b and c, we show the exciton dispersion in 1D along the M-K-Γ direction. In these plots, beside doing area averaging as mentioned above, we also integrate electron counts at constant radial distance r from the center of the exciton. To account for the two high symmetry directions, K-Γ and K-M direction, at the K-valley, we divided our data into 6 sectors as illustrated in figure S7A, and sum up the sectors that represent the same directions separately. Figure S7B shows the EDCs obtained by this procedure with peak position marked, these same peak positions are also shown in Fig. 3b.

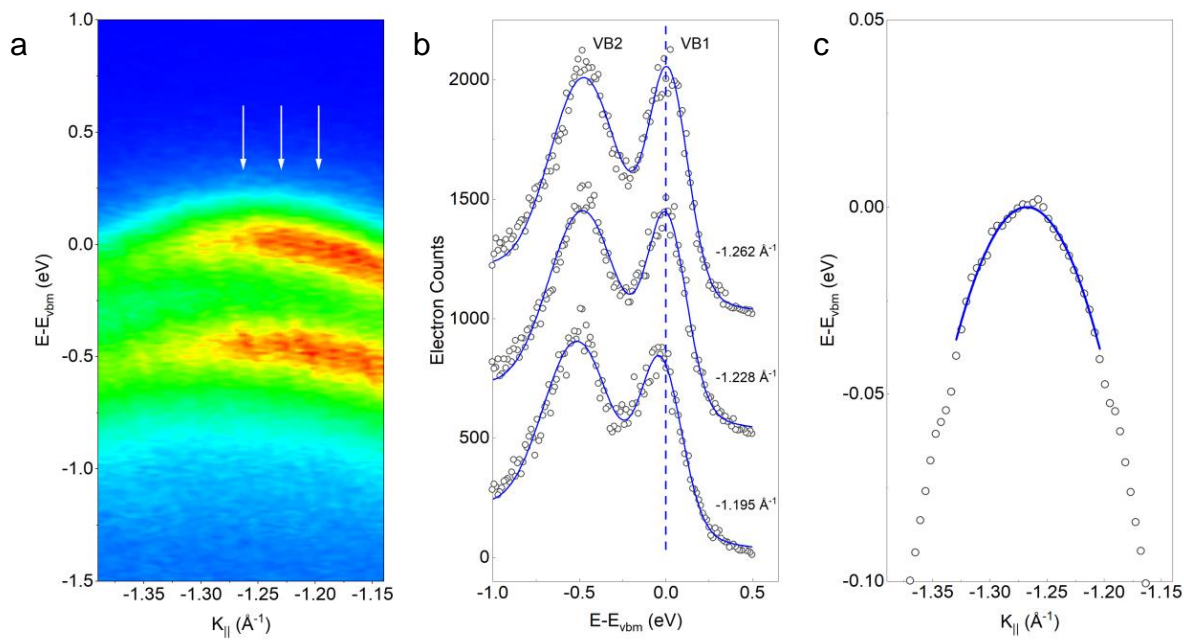

**Fig.S5. A,** Photoemission spectra of VB measured along the K-Γ direction at the K valley. (b), energy distribution curve (EDC) taken at positions indicated by the white arrows in (a), (open circles) are experiment data and (solid blue lines) are double Gaussian peak fitting to fit VB1 and VB2 positions. c, shows the position of VB1 obtained by peak fitting and the blue curve is a dispersion curve with a hole effective mass of -0.41 $m_e$ (bare electron mass).

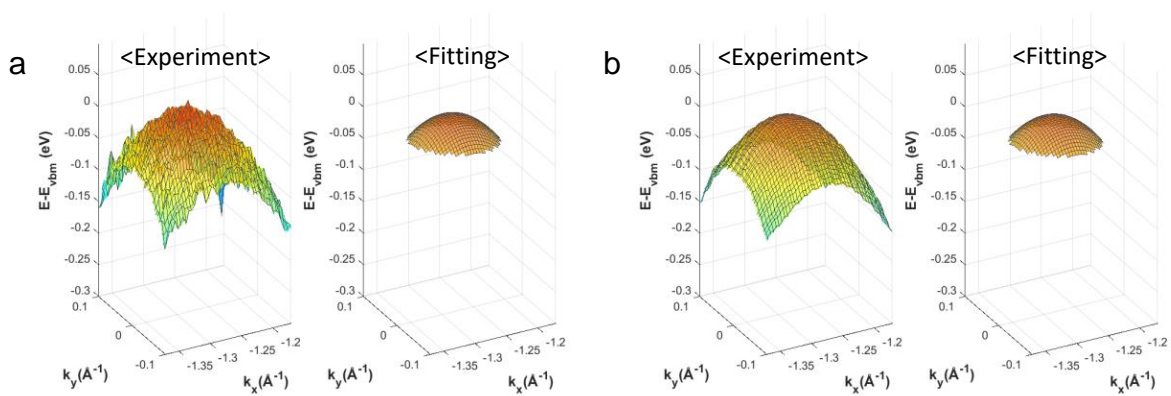

**Fig.S6.** 2D dispersion fitting of VB at the K valley. In (a), it is done with raw data without any intensity averaging. In (b), we first apply a 9x9 pixels rolling averaging on the raw data, corresponding to averaging an area of 0.04 x 0.04 Å$^{-2}$. In both cases, we obtain similar effective mass from the fitting.

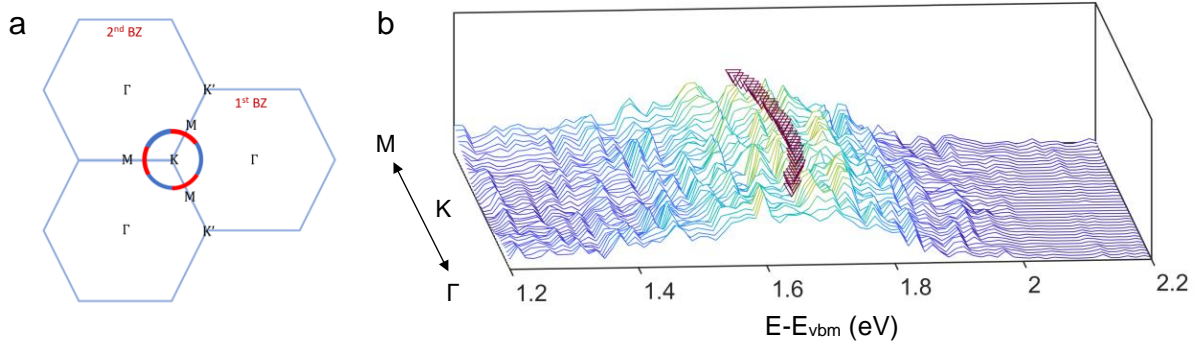

**Fig.S7.** 1D E(**k**) dispersion fitting of the K-valley exciton. (a) shows a schematic of the 1st and 2nd Brillouin zone. For data averaging, we divide data at radial distance r from the center of the K-valley into six sectors, the blue sectors are in the K-Γ direction and red sectors are in the K-M direction. (b) shows a waterfall plot for the EDCs obtained after averaging data in the 'sectors'. ▽ symbol marks the peak position obtained by peak fitting in each EDC.

Wavefunction in real space

In TR-ARPES, we measure probability density $I(\mathbf{k})$, that is the square of the excitonic wavefunction $\Psi(\mathbf{k})$ in momentum space. To obtain real space wavefunction, we first take $\sqrt{I(\mathbf{k})}$ of our data to obtain the wavefunction in momentum space. Assuming the phase of the wavefunction is flat for the 1s ground state of the exciton, we apply Fourier transform of $\Psi(\mathbf{k})$ and obtain $\Psi(\mathbf{r})$, the wavefunction of the exciton in real space. In Figs. 2C, F and G, we plotted the squared of the real space wavefunction $|\Psi(\mathbf{r})|^2$, which corresponds to the probability density of the electron relative to the hole in the exciton.

Estimation of the exciton densities

We estimate the exciton densities directly from the measured angle resolved photoemission intensity. Exciton signals is contained with an area A of 19.6 x $10^{-3}$ Å$^{-2}$ centered at the K valley. For the same area in the valence band, the corresponding density of states in the two spin split bands is around 1.4 x 10$^{13}$ electrons/cm$^2$. By extracting the photoelectron counts within the same area in k-space for the excitons and the valence bands, we can estimate the density of excitons $D_X$ with the following equation:-

$$D_X = \frac{C_X}{C_{VB}} \frac{|M_{VB}|^2}{|M_{CB}|^2} D_{VB}$$

where $D_{VB}$ is the density of state in the valence band, $C_X$ ($C_{VB}$) and $M_X$ ($M_{VB}$) are the measured photoelectron counts and photoemission matrix elements for the exciton states (valence bands). $M_X$ is approximately equal to photoemission matrix elements of conduction band electrons at K ($M_{CB}$). The ratio $\left|\frac{M_{VB}}{M_{CB}}\right|^2 \approx 1.5$ obtained from the first-principles calculations. With this method, we find exciton densities of 1.02 x10$^{11}$ cm$^{-2}$ for the 1.72 eV excitation.

Theory of exciton Dispersion

Here we describe in more detail the predicted anomalous dispersion signature of excitons in a photoemission experiment (Fig. 1c). For simplicity, let's assume excitons to have reached a quasi-equilibrium with a low enough temperature so the center of mass momentum Q is ~0. As described in text, for such an exciton $X$, the wavefunction $|\phi_X\rangle = \sum_{vc\mathbf{k}} A^X_{vc\mathbf{k}} c^\dagger_{c\mathbf{k}} c_{v\mathbf{k}} |0\rangle$ describes the electron-hole bound state, where $|0\rangle$ is the system ground state and $c_{i\mathbf{k}}$ destroys an electron with band index $i$ and wavevector $\mathbf{k}$. The factors $A^X_{vc\mathbf{k}}$ are then expansion coefficients of the exciton in terms of free electron-hole interband pairs, which can be interpreted as the envelope function of the Wannier exciton wavefunction in reciprocal space The initial energy of the system, prior to photoemission, is simply given by

$$E_{Initial} = h\nu + E_X \qquad (1)$$

where $h\nu$ is the energy of the XUV photon and $E_X$ is the energy of the exciton measured relative to the top of the Valence Band and is given by $E_X = E_g - E_b$. Here, $E_g$ is the energy of the quasiparticle bandgap and $E_b$ is the binding energy of the exciton. On photoemitting an electron with momentum $k$ from the exciton, we are left with a hole of momentum $-k$ in the sample. As in most ARPES experiments, we have assumed here the 'sudden approximation' [39]. Thus, the final energy of the system is given by

$$E_{Final} = W + E_{em} + \frac{\hbar^2 k^2}{2m_h} \qquad (2)$$

where $W$ is the work function, $E_{em}$ is the measured kinetic energy of the photoemitted electron, and the last term represents the kinetic energy of the hole with momentum $-k$ left behind in the sample. By energy conservation, we can write the kinetic energy of the photoemitted electron as:

$$E_{em} = (h\nu - W) + E_g - E_b - \frac{\hbar^2 k^2}{2m_h} \qquad (3)$$

From the last term, the anomalous negative dispersion of the photoemitted electron is clear.

Lastly, this derivation has been done for a Q=0 exciton, but can be extended for the case of finite temperature and Q≠0 excitons by considering the average energy of the photoemitted electron at each k. In our inhomogeneously broadened samples, we have assumed that the peak corresponds to the average.